\documentclass[conference]{IEEEtran}
\IEEEoverridecommandlockouts
\usepackage{cite}
\usepackage{amsmath,amssymb,amsfonts,bm}
\usepackage{algorithmic}
\usepackage{graphicx}
\usepackage{textcomp}
\usepackage{xcolor}
\usepackage{array}
\usepackage[tight,footnotesize]{subfigure}
\usepackage{nth}
\usepackage{threeparttable}
\usepackage{tablefootnote}
\usepackage{url}
\usepackage{hhline}
\usepackage{multirow}
\usepackage{tikz}
\usepackage{stackengine}

\def\BibTeX{{\rm B\kern-.05em{\sc i\kern-.025em b}\kern-.08em
    T\kern-.1667em\lower.7ex\hbox{E}\kern-.125emX}}

\graphicspath{ {./Figures/} }

\renewcommand{\baselinestretch}{0.97}

\begin{document}
\title{The Effect of Coupling Memory and Block Length on Spatially Coupled Serially Concatenated Codes}
\author{\IEEEauthorblockN{Mojtaba Mahdavi, 
		Muhammad Umar Farooq, Liang Liu, Ove Edfors, Viktor \"Owall, and Michael Lentmaier}
	\IEEEauthorblockA{Department of Electrical and Information Technology (EIT), Lund University, Lund, Sweden \\
		Emails: \{mojtaba.mahdavi, muhammad.umar\_farooq, liang.liu, ove.edfors, viktor.owall, michael.lentmaier\}@eit.lth.se}
	\thanks{{The simulations were performed on resources
		provided by the Swedish National Infrastructure for Computing (SNIC) at the center for scientific and technical computing at Lund University (LUNARC).}}}
\maketitle

\begin{abstract} 
Spatially coupled serially concatenated codes (SC-SCCs) are a class of spatially coupled turbo-like codes, which have a close-to-capacity performance and low error floor. In this paper we investigate the impact of coupling memory, block length, decoding window size, and number of iterations on the performance, complexity, and latency of SC-SCCs. Several design tradeoffs are presented to see the relation between these parameters in a wide range. Also, our analysis provides design guidelines for SC-SCCs in different scenarios to make the code design independent of block length. As a result, block length and coupling memory can be exchanged flexibly without changing the latency and complexity. {Also, we observe that the performance of SC-SCCs is improved with respect to the uncoupled ensembles for a fixed latency and complexity.} 
\end{abstract}


\section{Introduction}  \label{Sec.Introduction}
It has been shown that spatial coupling improves the  decoding threshold of low-density parity-check (LDPC) codes \cite{Lentmaier_TIT2010}. More specifically, the threshold of an iterative belief propagation (BP) decoder saturates to the threshold of the optimal maximum-a-posteriori (MAP) decoder \cite{Kudekar_2011,Kumar_2014}. The concept of spatial coupling has been extended to turbo-like codes in \cite{Saeedeh_TIT2017}, where it has been proven that threshold saturation also occurs for this class of codes. The decoding of spatially coupled codes can be done efficiently using  window decoding \cite{Fettweis_ICC2016, Iscan_CommLetter2018,Lentmaier_TC2017}. An information-coupled version of the turbo codes from the LTE standard was proposed in \cite{Yang_TC2018}. 
On the other hand, it has been demonstrated in \cite{Saeedeh_TC2019}  that spatial coupling leads to a new tradeoff between error floor and waterfall performance of turbo-like codes. As a result, with spatial coupling, serially concatenated codes (SCCs) achieve better performance than parallel concatenated codes (PCCs) in both the waterfall and the error floor regions \cite{Saeedeh_TC2019}. For this reason, spatially coupled serially concatenated codes (SC-SCCs) are selected as the focus of this paper.

From the analysis in \cite{Saeedeh_TIT2017} it can be seen that the decoding thresholds can be improved by increasing the coupling memory. But since the required size of the decoding window increases with the coupling memory, this option may not look appealing from a latency perspective.  In this paper, we take another approach and propose some design criteria that allow us to increase the coupling memory without increasing latency or complexity and without any performance loss. To this end, we investigate the effect of block size, coupling memory, window size, and number of iterations on the performance, complexity, and latency of SC-SCCs. Then, based on this analysis, we introduce a setup which allows us to fix the latency and complexity and trade between block length and coupling memory. This enables a fair comparison between different coding scenarios in terms of performance, complexity, and latency. 
\begin{figure}
	\centering
	\includegraphics[width=3.5in]{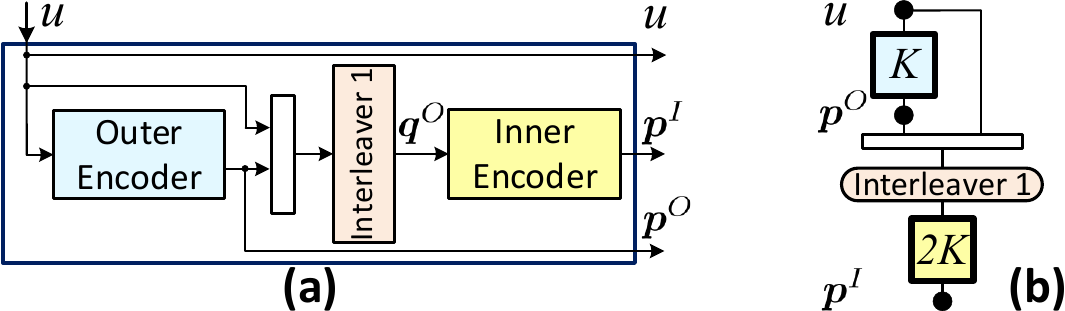}\\ \vspace{-9 pt}
	\caption{(a) SCC encoder structure, (b) Compact graph representation of SCC.} 
	\label{Fig.SCC_Encoder}\vspace{-7 pt}
\end{figure} 
\begin{figure*}[t]
	\centering 
	\includegraphics[width=7.2in]{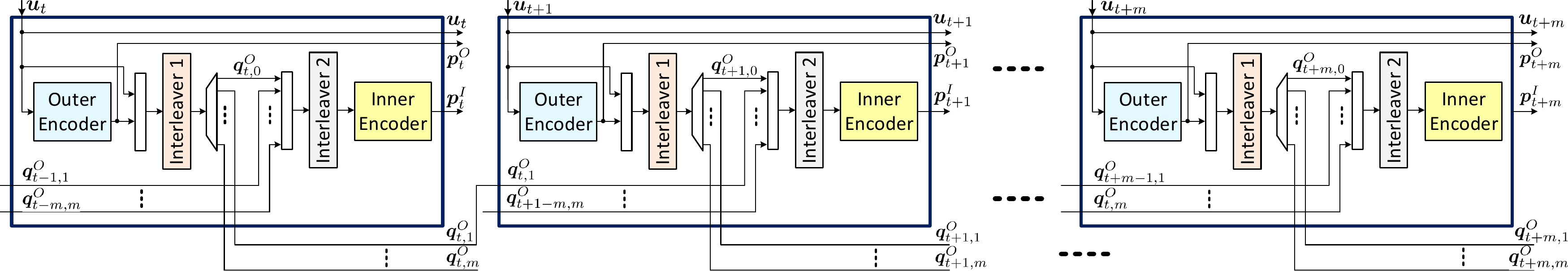}\\ \vspace{-9 pt}
	\caption{Structure of SC-SCC encoder with couping memory of $m$, which is built by spatial coupling of $m+1$ samples of SCC component encoders together.} 
	\label{Fig.SCSCC_Encoder}\vspace{-7 pt}
\end{figure*} 
As our approach lets us to flexibly exchange the block length with the coupling memory, spatial coupling allows a code designer to choose the strength and performance independently for a given block length. A performance loss for small block lengths is avoided in our scheme with continuous encoding and decoding. 

%

\section{Background}  \label{Sec.Background}
\subsection{SCC Encoder}  \label{Sec.SCC_Encoder}
The structure of an SCC component encoder is depicted in Fig.~\ref{Fig.SCC_Encoder}(a), which is made up of two recursive systematic convolutional (RSC) encoders concatenated in a serial manner using the interleaver. The left and right RSC encoders are called the outer and inner encoders with the trellis length of $K$ and $2K$, respectively. 
As shown in Fig.~\ref{Fig.SCC_Encoder}(a) the outer encoder receives the information sequence, $\bm{u}$, of length $K$ bits and produces the $K$-bit parity sequence $\bm{p}^O$. Then, the sequences $\bm{u}$ and $\bm{p}^O$ are multiplexed and permuted to generate the $2K$-bit sequence $\bm{q}^O$. This sequence is encoded by the inner encoder to produce the $2K$-bits parity sequence $\bm{p}^I$. Finally, the output of the SCC encoder is $\bm{v}=(\bm{u}, \bm{p}^O, \bm{p}^I)$. \par

\subsection{SC-SCC Encoder}  \label{Sec.SCSCC_Encoder}
We have built the SC-SCC encoder by coupling $m+1$ samples of SCC component encoders as shown in Fig.~\ref{Fig.SCSCC_Encoder}, where $m$ is the  coupling memory. Let us consider the encoding process at time instant $t$ to see how the inner and outer encoders are coupled together. As shown in Fig.~\ref{Fig.SCSCC_Encoder} the outer encoder receives the information bits, $\bm{u}_t$, and generates the parity sequence $\bm{p}^O_t$. Then, the pair of $(\bm{u}_t$, $\bm{p}^O_t)$ is permuted using \textit{Interleaver 1} to create a $2K$-bit sequence, $\bm{q}^O_t$. This sequence is divided into $m+1$ parts of equal size, which are named as $\bm{q}^O_{t,0}, \bm{q}^O_{t,1}, \ldots, \bm{q}^O_{t,m}$. This implies that $m+1$ should be smaller than $2K$ and also divide $2K$. The first subsequence, $\bm{q}^O_{t,0}$, is used to generate the input of the current inner encoder at time $t$ and the other ones, $\bm{q}^O_{t,1}, \ldots, \bm{q}^O_{t,m}$, will be used in the next inner encoders at time $t+1, \ldots, t+m$, respectively. Thus, at time $t$, the sequence $(\bm{q}^O_{t,0}, \bm{q}^O_{t-1,1}, \ldots, \bm{q}^O_{t-m,m})$, which is generated by the current and previous $m$ outer encoders is permuted by \textit{Interleaver 2} and sent to the inner encoder to produce the parity sequence $\bm{p}_t^I$. Finally, the output of the SC-SCC encoder at time $t$ is $\bm{v}_t=(\bm{u}_t, \bm{p}_t^O, \bm{p}_t^I)$. In this paper, a code rate of 1/3 is considered. Therefore, the output of the inner encoder, $\bm{p}_t^I$, is punctured such that only half of it, i.e. $K$ bits, is transmitted. 


Similar to protograph-based LDPC codes, we can describe turbo-like codes by compact graphs \cite{Saeedeh_TC2019}. The compact graph representation of SCCs is shown in Fig.~\ref{Fig.SCC_Encoder}(b), where the input and parity sequences are shown by black circles and referred to as \textit{variable nodes}. Also, the outer and inner code trellises are represented by squares, which are referred to as \textit{factor nodes} and labeled by the corresponding trellis lengths.\par  

The corresponding compact graph representation of an SC-SCC with coupling memory $m=1$ is shown in Fig.~\ref{Fig.Window_Decoding_Comapct_Graph}. The double circles are referred to as \textit{state variable nodes}, which transfer the encoder state at time  $t$ to the encoder at time $t+1$. As a result, our scheme performs the encoding continuously without termination. The reason behind this strategy is described in Section~\ref{Sec.Continuous_Encoding}. In a similar way, the compact graph of an SC-SCC with larger $m$ can be obtained. \par

\subsection{SC-SCC Window Decoder}  \label{Sec.SCSCC_Decoder}
\begin{figure}
	\centering
	\includegraphics[width=3.5in]{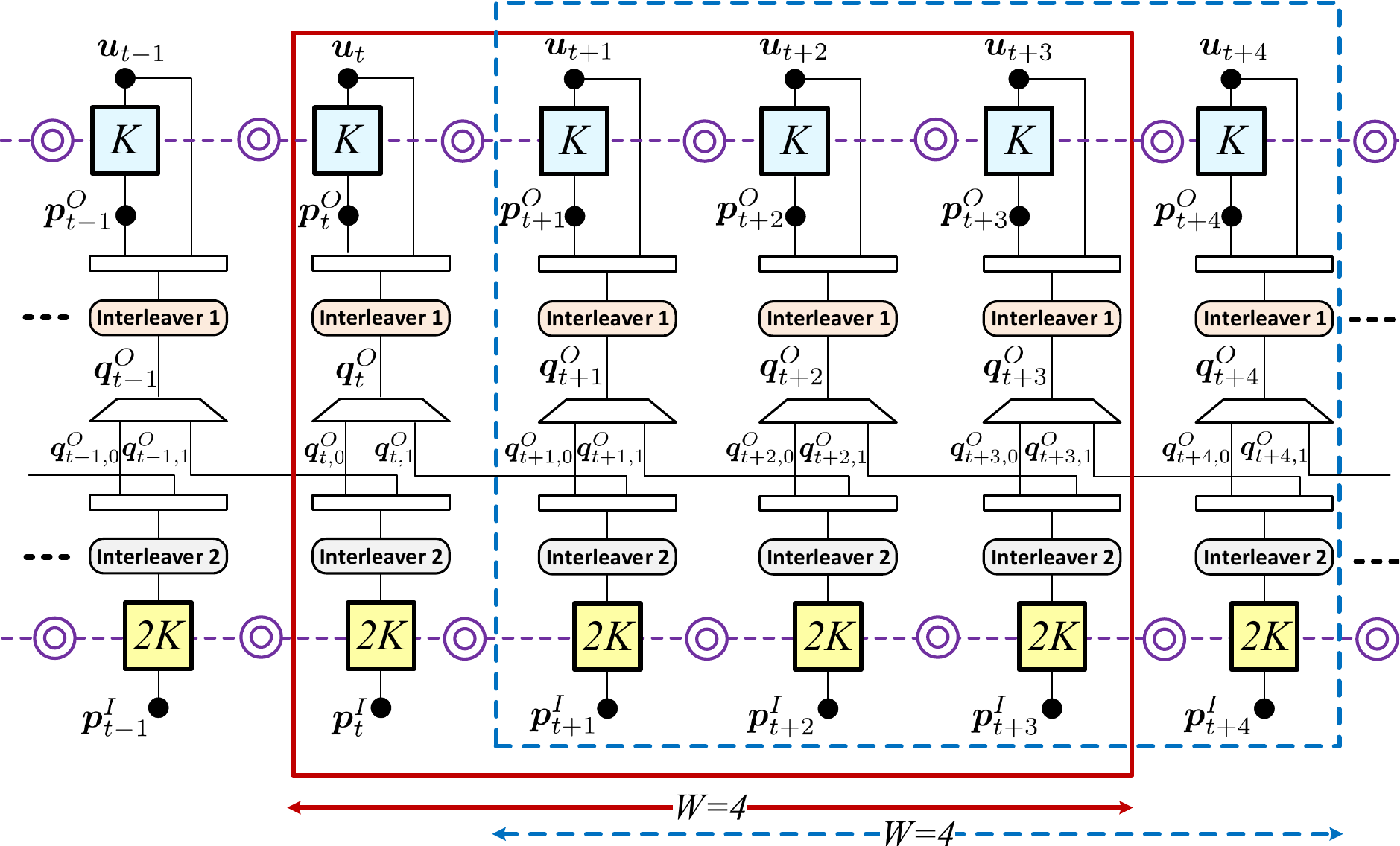}\\ \vspace{-9 pt}
	\caption{Compact graph representation of an infinite chain of SC-SCC for coupling memory $m$=1. Two decoding windows with $W$=4 blocks are shown.} 
	\label{Fig.Window_Decoding_Comapct_Graph}\vspace{-7 pt}
\end{figure}

{Analogously to LDPC codes, the nodes in an iterative message passing decoder exchange  log-likelihood ratios (LLRs) along the edges in  the graph (see Fig.~\ref{Fig.Window_Decoding_Comapct_Graph}).
	The inner and outer trellises are decoded using 
	the Bahl-Cocke-Jelinek-Raviv (BCJR) algorithm. Let us consider a decoding window of length $W$ blocks, which starts at time $t$ and ends at $t+W-1$, as shown by a solid rectangle in Fig.~\ref{Fig.Window_Decoding_Comapct_Graph}. Among these blocks, the first one to be decoded is referred to as the \textit{target block}, which is located to the leftmost side of the window.}\par  
{For all blocks with index $t^{\prime}=t,\dots,t+W-1$, first the inner and then the outer decoder perform $I_W$ decoding iterations as follows. In each iteration, the inner decoder receives three sequences: the channel LLR values $L_{ch}{(\bm{q}}^O_{t'})$ and $L_{ch}(\bm{p}^I_{t'})$, and the   a-priori LLR values, $L_{a}(\bm{p}^I_{t'})$, which are obtained based on the previous extrinsic LLRs of the corresponding outer decoder, $L_{e}(\bm{p}^O_{t'})$. The inner decoder produces the extrinsic LLRs, $L_{e}(\bm{p}^I_{t'})$, and sends them back to the outer decoder. Then, similarly, the outer decoder receives the channel LLR values $L_{ch}(\bm{u}_{t'})$ and $L_{ch}(\bm{p}^O_{t'})$, and the  a-priori LLRs, $L_{a}(\bm{p}^O_{t'})$, which are computed based on the previous extrinsic LLRs of the corresponding inner decoder, $L_{e}(\bm{p}^I_{t'})$. The outer decoder produces the extrinsic LLRs, $L_{e}(\bm{p}^O_{t'})$, and sends them back to the inner decoder.}
%
%
After $I_{W}$ iterations, the decoding of the target block, $\bm{u}_{t}$, is finished and the window is moved by one block. The same process is done for the next window, i.e. the dashed rectangle in Fig.~\ref{Fig.Window_Decoding_Comapct_Graph}, to decode the  target block $\bm{u}_{t+1}$. 

\textbf{Definitions:} The strength of spatially coupled codes depends on the \textit{constraint length}, which is defined as 
\begin{equation} \label{Eq:Constraint_Length}
\mathcal{C}=K\cdot(m+1),~~~~~~~~ m<W, 2K.
\end{equation}
Also, the \textit{structural latency} \cite{Rachinger_TC2015}, \cite{Rachinger_ISTC2014} is represented as  
\begin{equation} \label{Eq:Latency}
\mathcal{L}=W\cdot K,~~~~~~~ (\text{bit})
\end{equation}
which for simplicity we call latency in the rest of the paper.

\section{Design Guidelines for Flexible Choice of Block Size and Coupling Memory}  \label{Sec.Design_Guidelines}

\subsection{Using Higher Coupling Memory in a Fixed Latency}  \label{Sec.Fixed_Latency}
{A window decoder will perform very poorly if the window size, $W$, is smaller than $m+1$. }
 Thus, if a higher coupling memory is needed,  $W$ should be increased, which considerably increases the latency as stated in (\ref{Eq:Latency}). To solve this problem, we propose to reduce the block length, $K$, and increase the number of blocks per window, $W$, simultaneousely. As a result, a higher coupling memory can be used without changing the latency.\par 

\begin{figure}
	\centering
	\includegraphics[width=3.2in]{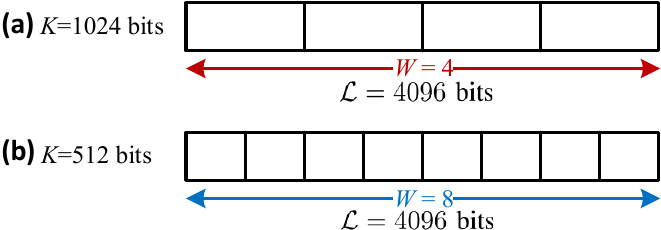}\\ \vspace{-9 pt}
	\caption{Two fixed-latency scenarios with different block length, $K$, and window size, $W$. (a) $K=1024$, $W=4$ and (b) $K=512$, $W=8$.} 
	\label{Fig.SCSCC_Decoder_Small_K}\vspace{-7 pt}
\end{figure}

Fig.~\ref{Fig.SCSCC_Decoder_Small_K} shows an example of an SC-SCC scheme with a latency of $\mathcal{L}=4096$ bits in two cases. In  Fig.~\ref{Fig.SCSCC_Decoder_Small_K}(a), four blocks of $K=1024$ bits per window are employed, which implies that the coupling memory cannot be larger than $m=3$. On the other hand, the same latency is achieved in Fig.~\ref{Fig.SCSCC_Decoder_Small_K}(b) by reducing the block length to $K=512$ bits and doubling the window size while the coupling memory can be increased up to $m=7$. Thus, depending on the block length, different window sizes should be considered to have a fixed latency and relax the limitation of the coupling memory. In Section~\ref{Sec.Coupling_Memory}, we will show that in a fixed latency scenario, a higher coupling memory results in a better performance compared to a smaller one. However, there are some challenges to employ small blocks and large coupling memory, which are addressed in our scheme as follows. 
 
\subsection{Continuous Encoding}  \label{Sec.Continuous_Encoding}
The classical way of encoding the SC-SCCs is to terminate the encoder after each block \cite{Saeedeh_TC2019}, i.e. encoder starts and ends in the zero state. The drawback of such schemes is a significant rate loss for small block lengths, $K$. One of the contributions of this paper is to perform  continuous encoding without termination after each block to avoid the rate loss specially for the small $K$. For this purpose, after encoding of the block at time $t$, the encoder state is passed to the encoder of the block at time $t+1$. Thus, the last state of the encoder at time $t$ is used as the starting state of the encoder at time $t+1$. To represent this concept, we have added the \textit{state variable nodes} to the SC-SCC graph, as shown by double circles in Fig.~\ref{Fig.Window_Decoding_Comapct_Graph}. 


\subsection{Performance Improvement of Boundaries}  \label{Sec.BCJR_Over_W}
\begin{figure}
	\centering
	\includegraphics[width=3.5in]{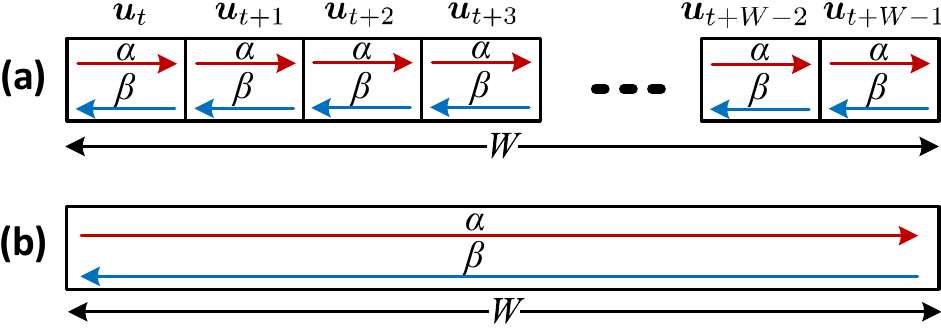}\\ \vspace{-9 pt}
	\caption{(a) Block-wise window decoding, (b) Proposed window decoding scheme. The window size, $W$, is the same in both cases.} 
	\label{Fig.SCSCC_Decoder_Win_BCJR}\vspace{-7 pt}
\end{figure}

The traditional window decoding algorithm usually works in a block-wise basis \cite{Lentmaier_TC2017} as shown in Fig.~\ref{Fig.SCSCC_Decoder_Win_BCJR}(a). Thus, at time instant $t$, the computation of $\alpha$ and $\beta$ are done in the forward and backward recursions for the block $\bm{u}_{t}$. Then, the results at time $t$ are used in decoding of the next block, $\bm{u}_{t+1}$. This method works properly for large block lengths, $K$. However, in case of small block lengths, running the BCJR for a very short trellis results in a poor performance at the boundaries between blocks. This is due to the unreliable states at the start and end of each trellis. Thus, the bits which are close to the boundaries will have a weak protection. To some extent this problem can be resolved by doing more iterations, but the computational complexity will be increased significantly. \par 

To address this challenge, we propose to perform the decoding over the whole window at once. As shown in Fig.~\ref{Fig.SCSCC_Decoder_Win_BCJR}(b), the $\alpha$ computation is done in the forward recursion over the whole window at once and then $\beta$ is computed in the backward recursion. Thus, regardless of the value of block length, $K$, and window size, $W$, in our scheme the BCJR algorithm is run one time per iteration over the whole window instead of $W$ times per iteration. As a result, the presented decoding scheme will be independent of the block length and window size. Also, since the trellis length becomes large the boundary states are more reliable, which can improve the performance especially for small block lengths. 

\subsection{Fixed Complexity}  \label{Sec.Fixed_Complexity}
Since, the trellis length of the inner decoder is twice the one of outer decoder, we define $2\mathcal{O}_D$ and $\mathcal{O}_D$ as the complexity of the inner and outer decoders. Due to the overlaps between the successive windows, shown in Fig.~\ref{Fig.Window_Decoding_Comapct_Graph}, each block is processed $W\cdot I_W$ times, where $I_{W}$ is the number of iterations per window position. 
%
 Thus, the computational complexity per bit is
\begin{equation}\label{Eq:Complexity_Per_Bit}
\mathcal{O}_{\text{bit}} = \frac{W\cdot(3\mathcal{O}_D)\cdot I_{W}}{K}  = \frac{3\mathcal{O}_D}{K} \cdot I_\text{eff} \  ,\vspace{-4pt}
\end{equation}
which is proportional to the {\em effective} number of iterations $I_\text{eff}=W\cdot I_W$, since $\mathcal{O}_{D}$ is proportional to $K$. Consequently, if the same $I_W$ is used for both cases in Fig.~\ref{Fig.SCSCC_Decoder_Small_K}(a) and (b), the scenario in Fig.~\ref{Fig.SCSCC_Decoder_Small_K}(b) will have higher complexity than the one in Fig.~\ref{Fig.SCSCC_Decoder_Small_K}(a), which is due to the larger $W$ and amount of overlaps between successive windows.\par 
  %
Here, $I_\text{eff}$ specifies how often the BCJR is run to decode a certain block. The goal is to adjust the $I_{W}$ such that the same $I_{\text{eff}}$ is achieved for all scenarios, which results in the same complexity per bit. This enables us to perform a fair comparison between different SC-SCC scenarios regardless of their block length, window size, and latency. For example, to have the same complexity in both scenarios in Fig.~\ref{Fig.SCSCC_Decoder_Small_K}, the $I_{W}$ in the second scenario, Fig.~\ref{Fig.SCSCC_Decoder_Small_K}(b), should be set to\vspace{-5pt}
\begin{equation}\label{Eq:Complexity_Iterations} 
I_{W_2} = \frac{W_1 \cdot I_{W_1}}{W_2},\vspace{-5pt}
\end{equation}
where $W_1$ and $I_{W_1}$ are corresponding to the case in Fig.~\ref{Fig.SCSCC_Decoder_Small_K}(a). It can be seen that less iterations per window, $I_W$, are used for smaller blocks. Also, it is important to point out that from a complexity perspective, both cases in Fig.~\ref{Fig.SCSCC_Decoder_Win_BCJR} are the same and it does not matter to run a long BCJR or several short ones.\par 

It is worth to mention that the computational complexity is not the only comparison metric that should be taken into account. There are other costs like the size of required memory and routing, which contribute to the hardware cost. However, these implementation issues are mainly related to the hardware architecture, which is not in the scope of this paper. 

\begin{table}[t]
	\caption{Different Scenarios of SC-SCCs with the Same Latency ($\mathcal{L}$), Constraint Length ($\mathcal{C}$), and Computational Complexity.}\vspace{-15pt}
	\begin{center}
		\begin{threeparttable}
			\begin{tabular}{|c|c|c|c|c|c|c|c|}
				\hline
				\multirow{2}{*}{$\mathcal{L}$\tnote{$\ddagger$}$~=16384$}&$K$&4096&2048&1024&512&256&128\\\cline{2-8}
				&$W$&4&8&16&32&64&128\\\cline{2-8}
				\multirow{2}{*}{$\mathcal{C}$\tnote{$\dagger$}$~=8192$}&$m$&1&3&7&15&31&63\\\cline{2-8}
				&$I_W$\tnote{$\ast$}&20&10&5&3\tnote{$\diamond$}&2\tnote{$\diamond$}&1\tnote{$\diamond$}\\\hline\hline	
				 
				\multirow{2}{*}{$\mathcal{L}=8192$}&$K$&2048&1024&512&256&128&64\\\cline{2-8}
				&$W$&4&8&16&32&64&128\\\cline{2-8}
				\multirow{2}{*}{$\mathcal{C}=4096$}&$m$&1&3&7&15&31&63\\\cline{2-8}
				&$I_W$&20&10&5&3\tnote{$\diamond$}&2\tnote{$\diamond$}&1\tnote{$\diamond$}\\\hline\hline
				
				\multirow{2}{*}{$\mathcal{L}=4096$}&$K$&1024&512&256&128&64&32\\\cline{2-8}
				&$W$&4&8&16&32&64&128\\\cline{2-8}
				\multirow{2}{*}{$\mathcal{C}=2048$}&$m$&1&3&7&15&31&63\\\cline{2-8}
				&$I_W$&20&10&5&3\tnote{$\diamond$}&2\tnote{$\diamond$}&1\tnote{$\diamond$}\\\hline\hline
				
				\multirow{2}{*}{$\mathcal{L}=2048$}&$K$&512&256&128&64&32&-\tnote{$\triangleleft$}\\\cline{2-8}
				&$W$&4&8&16&32&64&-\\\cline{2-8}
				$\mathcal{C}=1024$&$m$&1&3&7&15&31&-\\\cline{2-8}
				&$I_W$&20&10&5&3\tnote{$\diamond$}&2\tnote{$\diamond$}&-\\\hline\hline
				
				\multirow{2}{*}{$\mathcal{L}=1024$ }&$K$&256&128&64&32&16&-\tnote{$\triangleleft$}\\\cline{2-8}
				&$W$&4&8&16&32&64&-\\\cline{2-8}
				\multirow{2}{*}{$\mathcal{C}=512$}&$m$&1&3&7&15&31&-\\\cline{2-8}
				&$I_W$&20&10&5&3\tnote{$\diamond$}&2\tnote{$\diamond$}&-\\\hline\hline
			\end{tabular}
		\label{Tb.Fixed_Latency_Constraint}
			\begin{tablenotes}[para]
				\item[\hspace{-6pt}$\ddagger$] {Calculated using (\ref{Eq:Latency})}~~~\item[\hspace{-5pt}$\dagger$] {Calculated using (\ref{Eq:Constraint_Length})}~~\item[\hspace{-5pt}$\ast$] {Calculated using (\ref{Eq:Complexity_Iterations})}\\
				\item[\hspace{-6pt}$\diamond$] {Rounded to the nearest largest integer number. Also, $I_{\text{eff}}=80$ is used to have the same complexity and to perform $I_W\geq1$ for all scenarios.}\\
				\item[\hspace{-6pt}$\triangleleft$] {Not available in this scenario since (\ref{Eq:Latency}) implies that $m<2K$.}
			\end{tablenotes}
		\end{threeparttable} \vspace{-10pt}
	\end{center}
\end{table}

\section{Performance Evaluation}  \label{Sec.Performance_Evaluation}
We have investigated the effect of code properties (e.g. $K$, $m$) and also the decoding parameters (e.g. $W$, $I_{\text{eff}}$) on the performance and complexity of SC-SCCs. To this end, we have defined and used five SC-SCC scenarios as listed in Table~\ref{Tb.Fixed_Latency_Constraint}. In each scenario the latency $\mathcal{L}$, constraint length $\mathcal{C}$, and complexity are fixed, which are obtained by different combinations of $K$, $W$, $m$, and $I_W$ in a wide range. In the simulations, the information sequence is modulated using binary phase shift keying (BPSK) modulation and transmitted through the additive white Gaussian noise (AWGN) channel\footnote{We have picked a set of pseudo-random interleavers and fixed them for all the code sequences through the simulations. }.

\subsection{Effect of Coupling Memory on the Performance} \label{Sec.Coupling_Memory}
\begin{figure}
	\centering
	\includegraphics[width=3.5in]{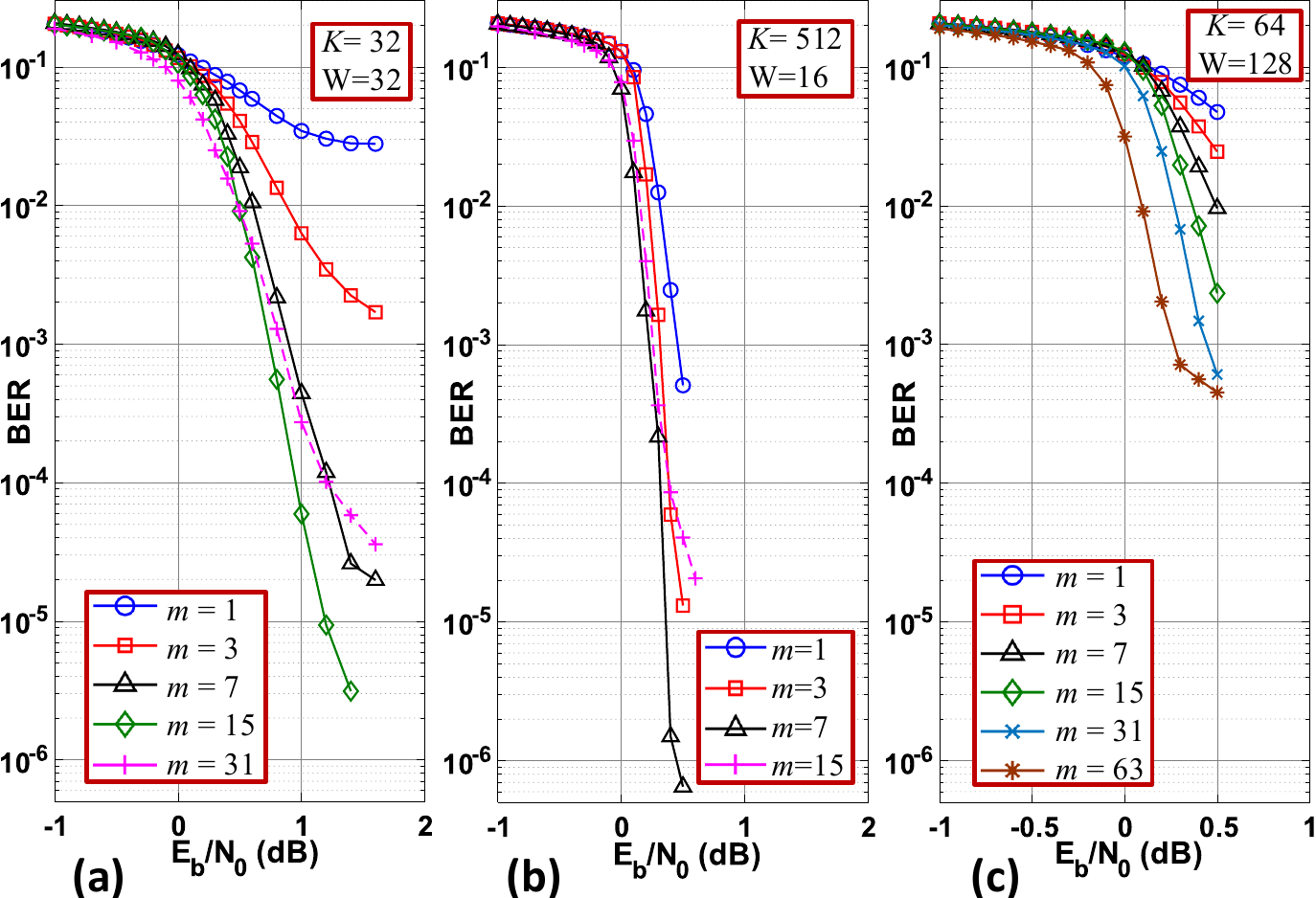}\\ \vspace{-9 pt}
	\caption{The effect of coupling memory, $m$, on the performance in different scenarios. The latency is (a) $\mathcal{L}=1024$, (b) $\mathcal{L}=8192$, and (c) $\mathcal{L}=8192$ bits. The complexity is the same for all scenarios ($I_{\text{eff}}=80$ for all cases).} 
	\label{Fig.BER_Effect_m}\vspace{-7 pt}
\end{figure}

We have investigated the effect of coupling memory, $m$, on the performance of SC-SCCs. The goal is to fix the window size, $W$, and block length, $K$, and then find the value of coupling memory, $m$, which leads to the best performance. 
As an example, this concept is investigated for three cases: \{$\mathcal{L}=1024$, $K=32$, $W=32$\}, \{$\mathcal{L}=8192$, $K=512$, $W=16$\}, and \{$\mathcal{L}=8192$, $K=64$, $W=128$\} and the corresponding results are depicted in Fig.~\ref{Fig.BER_Effect_m}(a)-(c). As a result, by increasing the coupling memory up to $m=W/2-1$ the performance will be improved considerably (i.e. 0.2 dB to 1.1 dB). Also, the error floor goes down to the lower BERs and the waterfall performance becomes better. 
But, if $m>W/2-1$ the performance will be degraded, as shown with doted curves in Fig.~\ref{Fig.BER_Effect_m}. This is due
to the fact that in such a case we cannot see even one constraint
length, $\mathcal{C}$, inside the window (see Fig.~\ref{Fig.Window_Decoding_Comapct_Graph}) and therefore the
performance of the decoder cannot fully exploit the code. Thus, for a given $K$ and $W$ the coupling memory of $m=W/2-1$ results in the best performance in such a setup. It is worth to mention that this performance improvement is achieved without compromising the latency and complexity.

\begin{figure*} 
	\centering 
	\includegraphics[width=7.2in]{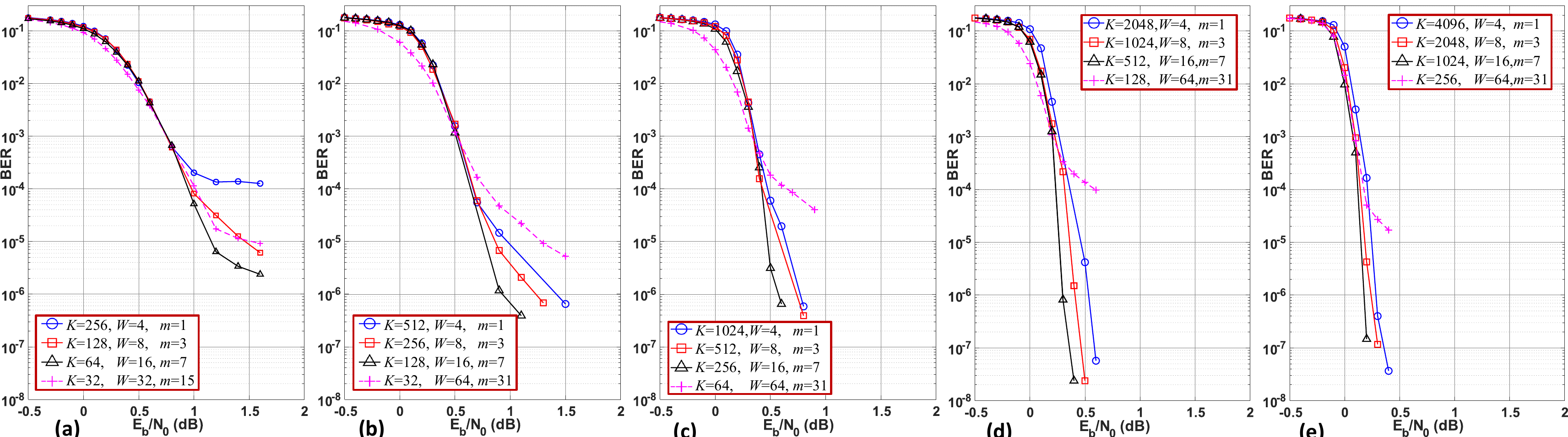}\\ \vspace{-9 pt}
	\caption{BER Performance of the scenarios in Table~\ref{Tb.Fixed_Latency_Constraint}, where the latency and constraint are fixed to (a) $\mathcal{L}=1024, \mathcal{C}=512$, (b) $\mathcal{L}=2048, \mathcal{C}=1024$, (c) $\mathcal{L}=4096, \mathcal{C}=2048$, (d) $\mathcal{L}=8192, \mathcal{C}=4096$, and (e) $\mathcal{L}=16384, \mathcal{C}=8192$. The same complexity is considered for all scenarios by choosing $I_{\text{eff}}=80$.}
	\label{Fig.BER_Total}\vspace{-7 pt}
\end{figure*} 
 
\subsection{Shorter Block Length with Higher Coupling Memory}\label{Sec.Shorter_Block_Length}
As mentioned in Section~\ref{Sec.Coupling_Memory}, the coupling memory of $m=W/2-1$ leads to the best performance for the given $K$ and $W$. This choice of coupling memory results in the constraint length of $\mathcal{C}=K\cdot(m+1)=K\cdot W/2$, which can be achieved by either a small $K$ and large $m$ or a large $K$ and small $m$ while the latency remains fixed ($\mathcal{L}=2\mathcal{C}$). For example, $\{K=256, W=4, m=1\}$ and $\{K=64, W=16, m=7\}$ achieve the same latency of $\mathcal{L}=1024$ bits and constraint length of $\mathcal{C}=512$ as shown in Table~\ref{Tb.Fixed_Latency_Constraint}. \par

It is expected to achieve same performance for the scenarios, which have the same constraint length. We have investigated this concept for the five scenarios in Table~\ref{Tb.Fixed_Latency_Constraint} and the simulation results are shown in Fig.~\ref{Fig.BER_Total}(a)-(e). In each scenario the latency and constraint length are fixed while the  complexity is the same for all of them by considering $I_{\text{eff}}=80$. Simulation results show that, for a certain latency and constraint length, selecting a small block length, $K$, and large coupling memory, $m$, can lead to a better performance compared to a large block length and small coupling memory. This performance improvement can be seen in both waterfall and error floor regions in Fig.~\ref{Fig.BER_Total}. \par

As a result, our analysis reveals the flexibility of SC-SCCs such that for a given latency and constraint length, it is possible to make the block length smaller and use higher coupling memory while we get the same or even better performance compared to the larger $K$. It is worth to mention that, in case of very small $K$ the performance degrades and an error floor appears at high BERs, which are shown by the dashed curves in Fig.~\ref{Fig.BER_Total}(a)-(e). This is mainly due to the fact that in our scheme, we have employed independent random interleavers to show how the performance changes for different block lengths. But, in case of very small $K$ the short-length random interleavers are not efficient. In such cases, the interleavers should not be designed independently\footnote{We  plan to investigate a joint interleaver design for small block lengths, $K$, in our future research.}. Moreover, since we want to have the same complexity for different block lengths, $I_W$ would be very low (e.g. $I_W=1,2,3$ in Table~\ref{Tb.Fixed_Latency_Constraint}) for the very small $K$, which degrades the performance. Note that a small or large block length is relative  to the latency. For example, $K=128$ is considered as a large $K$ in case of $\mathcal{L}=1024$, while it is a small $K$ for $\mathcal{L}=8192$. 


\subsection{Performance Comparison with Uncoupled Codes}
Fig.~\ref{Fig.BER_SCC} shows the performance comparison between the presented SC-SCC scheme and the uncoupled ensembles, SCC, for different latencies, $\mathcal{L}$, and block lengths, $K$. To have a fair comparison, the same complexity is considered for both SC-SCC and SCC regardless of $\mathcal{L}$ and $K$ as described in Section~\ref{Sec.Fixed_Complexity}. It can be seen that spatial coupling significantly improves the performance of the SCC and makes it much closer to the capacity. Having considered the same interleaver size (i.e. fixed block length, $K$) the SC-SCC achieves around 1 dB better performance than the corresponding SCC scheme with the same $K$. Also, in case of equal latency, the SC-SCC scheme still achieves around 0.5 dB better performance than the SCC at the BER of $10^{-4}$. The latency of an SCC is $\mathcal{L}=K$. 
Moreover, Fig.~\ref{Fig.BER_SCC} shows that the performance improvement resulting from increasing the latency is more pronounced in SC-SCCs than SCCs. More specifically by increasing the latency from $\mathcal{L}=1024$ to $\mathcal{L}=32768$ bits, 0.7 dB and 1.1 dB performance improvement is achieved in SCC and SC-SCC, respectively. \par 

It is worthwhile to mention that, even with lower latency, the SC-SCC can achieve even better performance than the SCC scheme. For example, as shown in Fig.~\ref{Fig.BER_SCC}, the SC-SCC with $\mathcal{L}=8192$ has better performance than the SCC with $\mathcal{L}=32768, 16384$. This means that by just increasing the latency and block length the SCC cannot achieve better performance than the SC-SCC, which is due to the threshold improvement resulting from spatial coupling.
{Moreover, asymptotic decoding thresholds of the SCC and SC-SCC  ensembles for the AWGN channel are depicted using vertical lines in Fig.~\ref{Fig.BER_SCC}. These values are computed using the erasure channel prediction method introduced in \cite{Umar_ISIT_2018}.
%



\begin{figure}
	\centering
	\includegraphics[width=3.5in]{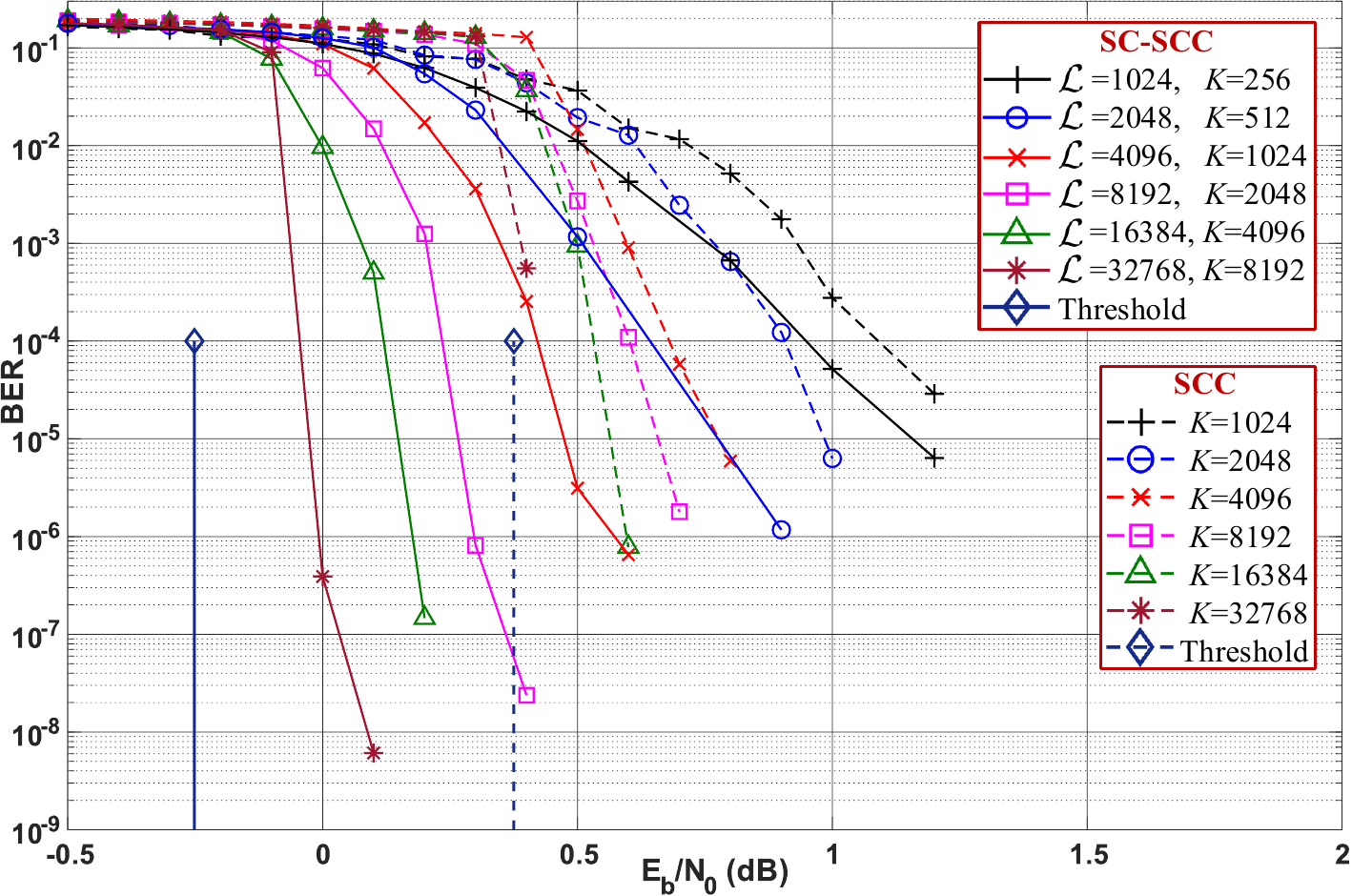}\\ \vspace{-10 pt}
	\caption{Performance comparison between proposed SC-SCC and SCC for different block lengths, $K$, and latencies, $\mathcal{L}$.  The same complexity is considered for all cases by choosing $I_{\text{eff}}=80$.} 
	\label{Fig.BER_SCC}\vspace{-9 pt}
\end{figure}

\section{Design Tradeoffs}  \label{Sec.Design_Tradeoffs}
So far we assumed fixed latency and complexity in the evaluations. {Now we are interested to see the performance gain of SC-SCCs if we increase the latency or the complexity.}

\subsection{Performance-Latency Tradeoff}\label{Sec.Performance_Latency_Tradeoffs}
As mentioned in Section~\ref{Sec.Coupling_Memory}, for a given latency, $\mathcal{L}$, and complexity, the performance can be improved by increasing the constraint length $\mathcal{C}$, i.e., larger $m$ or $K$. Now, we want to see if for a given constraint length and complexity, it is possible to improve the performance by increasing the latency? More specifically, having considered a fixed $K$ and $m$, how does the performance change if we make the window size, $W$, larger? 
We have investigated this concept and the results are shown in Fig.~\ref{Fig.BER_Effect_W}. For a given constraint length and complexity, if we only change the window size (e.g. doubling $W$) we will not gain too much in performance while the latency is increased (e.g. twice latency). 
So, if the targeted application can tolerate the higher latencies, it is better to increase the coupling memory, $m$, as well to make the code stronger rather than just increasing the window size, $W$. As a result, this strategy provides the effective use of a certain latency to achieve better performance. It is worth to mention that, the larger window size enables us to employ a higher coupling memory. This is due to the limitations on the coupling memory, i.e. $m\le~W/2-1$, as described in Section~\ref{Sec.Coupling_Memory}.\par 

\begin{figure}
	\centering
	\includegraphics[width=3.5in]{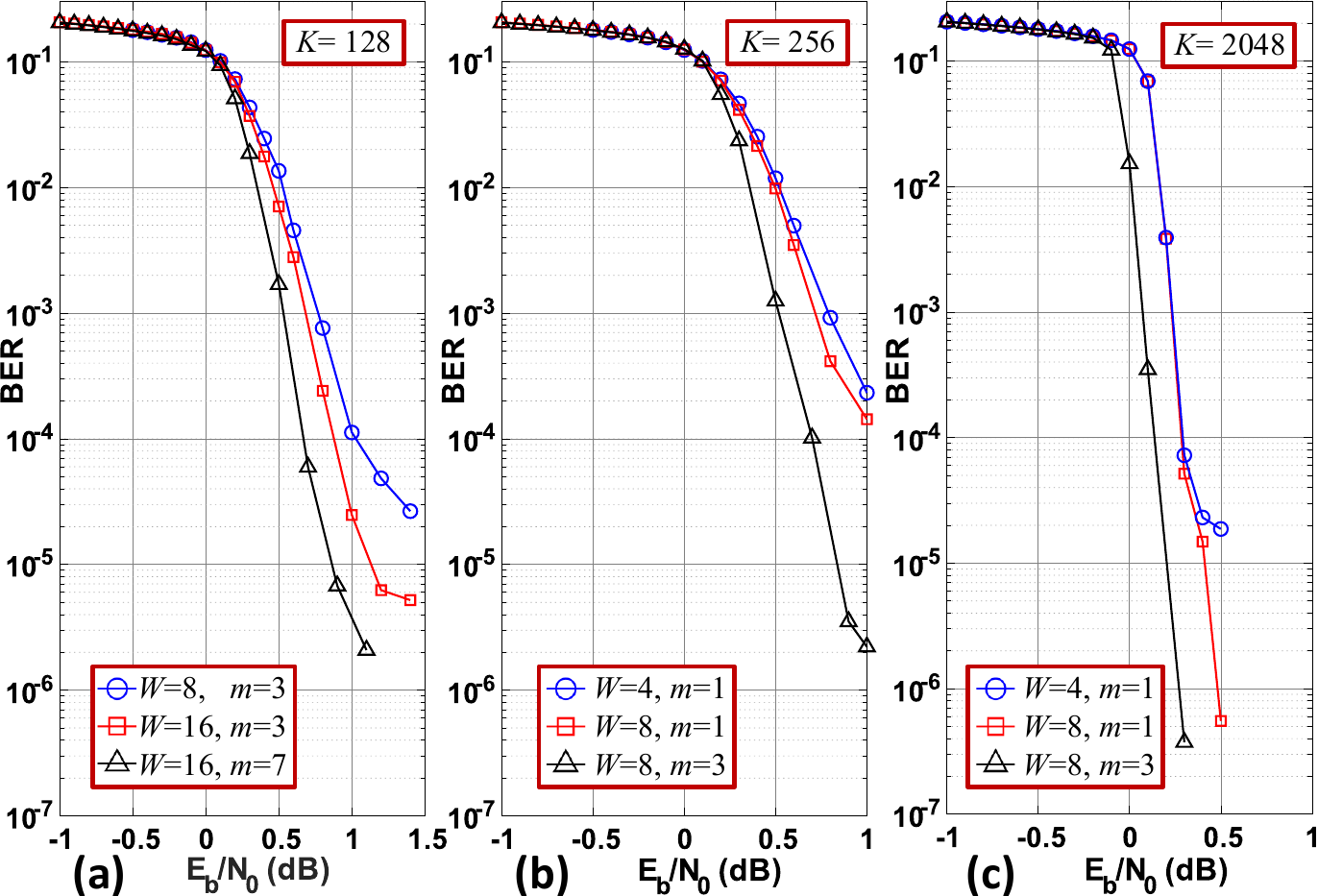}\\ \vspace{-9 pt}
	\caption{Simulation results to investigate the effect of window size, $W$, on the performance. The block length, $K$, is (a) 128 bits, (b) 256 bits, and (c) 2048 bits. The computational complexity is the same for all scenarios ($I_{\text{eff}}=80$).} 
	\label{Fig.BER_Effect_W}\vspace{-7 pt}
\end{figure}

\begin{figure}
	\centering
	\includegraphics[width=3.5in]{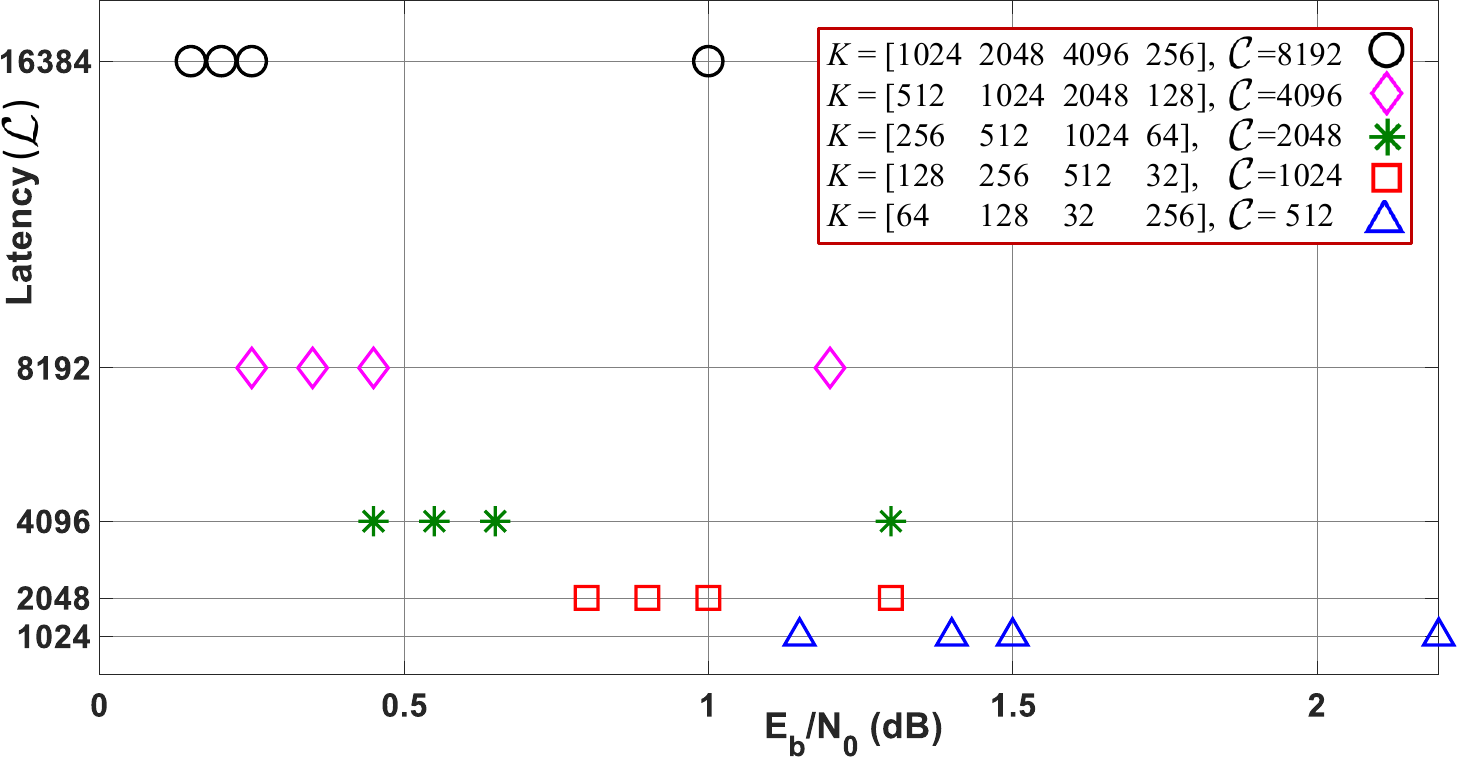}\\ \vspace{-9 pt}
	\caption{The latency-performance tradeoff for the scenarios in Table~\ref{Tb.Fixed_Latency_Constraint}. The required $\text{E}_\text{b}/\text{N}_\text{0}$ to achieve BER of $10^{-5}$ is shown in x-axis. The listed values of $K$ in the legend are corresponding to the markers from left to right. The same computational complexity is considered for all scenarios ($I_{\text{eff}}=80$).}
	\label{Fig.SNR_Latency_Tradeoff}\vspace{-7 pt}
\end{figure}

Fig.~\ref{Fig.SNR_Latency_Tradeoff} shows the tradeoff between latency and performance for the scenarios in Table~\ref{Tb.Fixed_Latency_Constraint}. In each scenario the latency, $\mathcal{L}$, and constraint length, $\mathcal{C}$, are fixed and the complexity is the same for all scenarios. In this figure, x-axis shows the required $\text{E}_\text{b}/\text{N}_\text{0}$ to achieve the BER of $10^{-5}$ in all scenarios. The markers, which tend to the lower left corner of Fig.~\ref{Fig.SNR_Latency_Tradeoff} are corresponding to the scenarios with a low latency, $\mathcal{L}$, and good performance. Thus, by considering this tradeoff the proper values of $W$, $K$, and $m$ can be obtained. 


\subsection{Performance-Complexity Tradeoff}
It is worthwhile to see the effect of number of iterations on the performance: how will the performance be improved if we would spend more complexity? We have considered different effective number of iterations, $I_{\text{eff}}$, for the SC-SCC scheme in case of $\mathcal{L}=4096$ and $\mathcal{L}=16384$ bits latencies. The corresponding simulation results are shown in Fig.~\ref{Fig.BER_Different_Iter}(a) and (b), respectively. Simulation results show that the performance improvement resulting from a higher effective number of iterations, $I_{\text{eff}}$, is more pronounced in the high latency, $\mathcal{L}$, scenarios than the low latency scenarios. Also, it can be seen that by spending more complexity the error floor goes down and it happens at a much lower BER.

\begin{figure}
	\centering
	\includegraphics[width=3.5in]{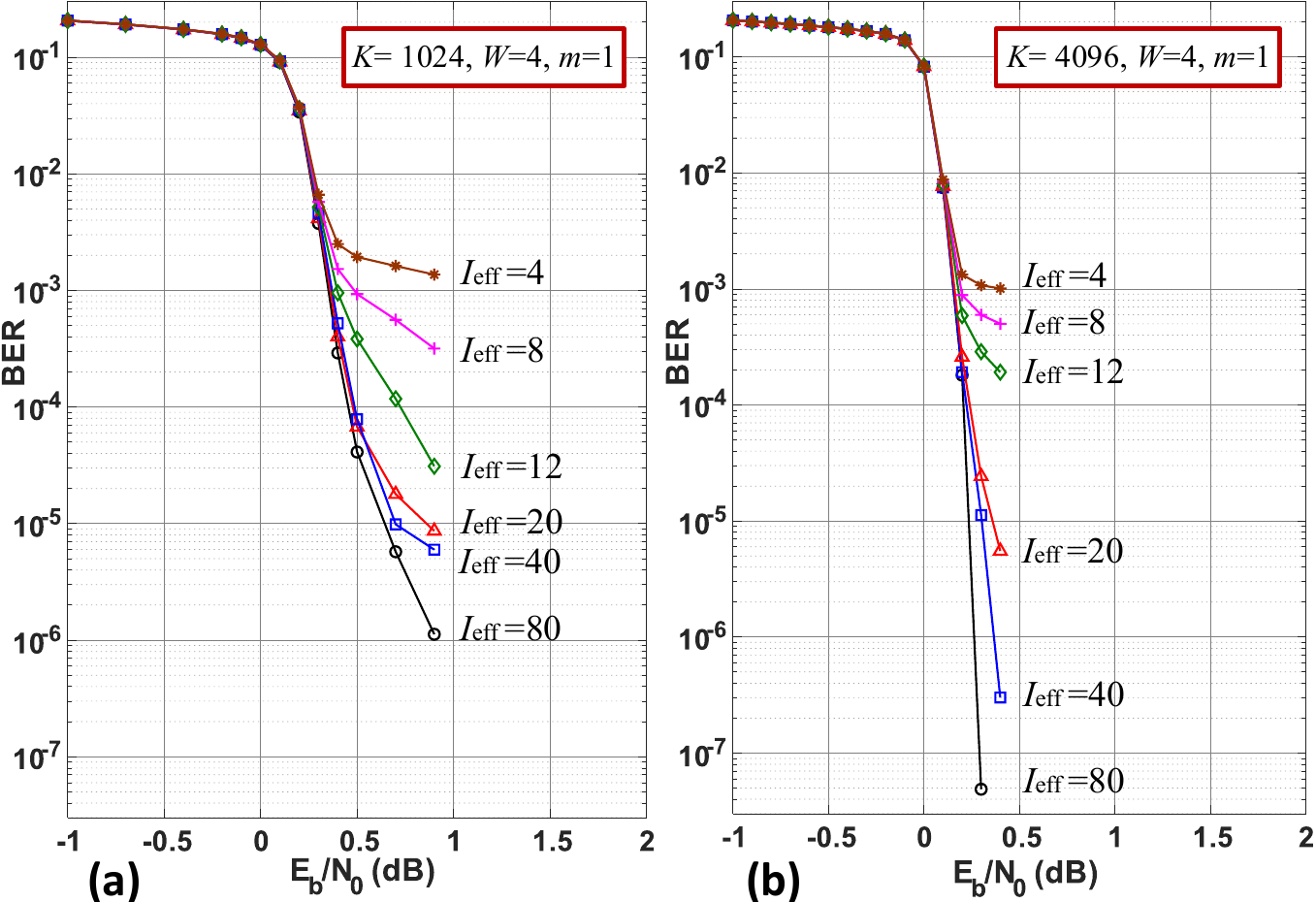}\\ \vspace{-9 pt}
	\caption{Simulation results to investigate the effect of number of iterations on the performance. The latency equals (a) $\mathcal{L}=4096$ and (b) $\mathcal{L}$=16384 bits.} 
	\label{Fig.BER_Different_Iter}\vspace{-7 pt}
\end{figure}

\section{Conclusion} \label{Conclusion}  
We have investigated the effect of coupling memory, block length, window size, and number of iterations on the performance, complexity, and latency of SC-SCCs. Our approach provides the flexibly to exchange the block size with the coupling memory, which makes the code design independent of the block length. We have demonstrated that how the higher coupling memory can be used without increasing the latency and complexity. Moreover, we have shown that SC-SCCs can achieve better performance than the uncoupled ensembles with the same latency and complexity.  




\end{document}